\documentclass[12pt,oneside]{article}
\usepackage[english]{babel}
\usepackage{amsmath}
\usepackage{amsfonts}
\usepackage{amssymb, pst-tree}

\usepackage{tikz}

\usepackage{theorem}
\usepackage{sgame}

\newtheorem{theorem}{Theorem}

\newtheorem{corollary}{Corollary}

\newtheorem{proposition}{Proposition}
\newtheorem{lemma}{Lemma}
{\theorembodyfont{\rmfamily}
\newtheorem{example}{Example}

} {\theorembodyfont{\rmfamily}

}
\newenvironment{proof}[1][]
   {\par\medbreak{\noindent\bfseries Proof#1\quad}}
   {\hbox{}\hfill $\square$\bigbreak}

\DeclareMathOperator*{\ptop}{top}
\begin{document}

\title{Roommates with Convex Preferences}

\author{\textsc{Sophie Bade}\footnote{Royal Holloway College, Egham Hill TW10 0EX, UK,  sophie.bade@rhul.ac.uk. Funding through the ARCHES prize by the German Minerva foundation is gratefully appreciated.}}

 \maketitle

\begin{abstract}
Roommate problems with convex preferences always have stable matchings.  Efficiency and individual rationality are, moreover, compatible with strategyproofness in such convex roommate problems. Both of these results fail without the assumption of convexity. In the environment under study, preferences are convex if and only if they are single peaked. Any individually rational and convex roommate problem is homomorphic to a marriage market where an agent's gender corresponds to the direction of the agent's top-ranked partner. The existence of stable matchings then follows from the existence of stable matchings in marriage markets. To prove the second existence result, I define an efficient, individually rational, and strategyproof mechanism for convex roommate problems.
To calculate outcomes, this mechanism starts with all agents being single and then gradually reassigns agents to better partners by performing minimal Pareto improvements. Whenever it becomes clear that some agent cannot be part of any further Pareto improvement, such an agent is matched.

\medskip

\end{abstract}
\newpage

\section{Introduction}

Convexity drives many results in economics and its power extends beyond models in Euclidean spaces: Richter and Rubinstein \cite{RiRu}, for example, prove a version of the  second fundamental theorem of welfare economics that uses a notion
of convexity for consumption spaces without an algebraic structure. One particular notion of convex preferences over finitely many objects has a long pedigree in economics:
When objects are ordered on a line, then a linear order is convex if and only if it is single-peaked.
In the context of social choice, single peakedness makes all the difference: While Gibbard \cite{Gibbard} and Satterthewaite \cite{Satterthwaite} show that generic domains permit no strategyproof and efficient social choice rules, the median rule satisfies both these properties on the domain of single-peaked preferences.

Matching is no different: non-convexities lie at the root of many impossibility results.
The present paper shows that two impossibility results for (standard) roommate problems do not extend to the case where all agents have convex preferences. While generic roommate problems may lack stable matchings, Corollary \ref{theorem: stability} shows that   roommate problems with convex/single-peaked preferences always have stable matchings.
However, even when all agents have convex preferences, stability clashes with incentive compatibility (Proposition \ref{proposition: 3 agents stable} - part a). But there are incentive-compatible mechanisms when we weaken stability to a combination of efficiency and individual rationality: Theorem \ref{theorem: MC satisfies the axioms} constructs a strategyproof mechanism for convex roommate problems, that is efficient and individually rational. Without the restriction to convex preferences no such mechanism exists (Proposition \ref{proposition: 3 agents stable} - part b).

 In a roommate problem, each agent must either find a roommate or stay single. Each agent has a strict preference over all possible roommates, including herself. If an agent ranks herself above a different agent, then she prefers staying single to being matched with that agent. A matching assigns to each agent a partner (possibly herself). A mechanism maps the set of all preferences profiles to the set of all matchings. It
is strategy-proof if no agent is ever better off by lying about his preference. Individually rational mechanisms never match an agent with a partner she deems worse than being single. Efficient mechanisms map each profile of preferences to a matching that is efficient at that profile. A matching is stable if it is individually rational and if no two agents prefer each other to their assigned partners.

 To illustrate the assumption of single peakedness, say  that all agents can be ranked from  quiet bookworm to wild party animal. Each agent then has a view on his ideal roommate. Some agent may just want to find a roommate that is as similar to himself as possible. Other agents may hope to change:  a social butterfly may want to match with a reasonably studious roommate in the hope that the roommate's attitude to learning may rub off. Conversely, a shy nerd may hope to increase their social life by matching with the perfect dinner party hostess. Single peakedness entails that the agents do not just have an idea of their ideal roommate, their utility strictly decreases as they move away from their ideal partner.

\section{Literature}

Gale and Shapley \cite{GaleShapley} introduced roommate problems  as a contrast to marriage markets. Marriage markets are roommate problems with two genders, where no two agents of the same gender may form a pair. Gale and Shapley's \cite{GaleShapley} seminal contribution lies in showing that  marriage markets always have stable matchings. To show that the  partition into two genders is crucial, Gale and Shapley \cite{GaleShapley} construct a roommate problem without a stable matching.

Lemma \ref{lemma: embedding} shows that individually rational convex roommate problems are marriage markets with convex preferences. Using the equivalence established in
Lemma \ref{lemma: embedding}, Corollary \ref{theorem: stability} imports the existence of stable matchings from Gale and Shapley \cite{GaleShapley} into convex roommate problems.  Corollary \ref{theorem: stability} was already shown by Chung \cite{Chung}. Chung  \cite{Chung} develops a necessary condition for the existence of stable matchings in roommate problems called ``no odd rings'' and shows that convex roommate problems do satisfy this condition.

Proposition \ref{proposition: stable} strengthens
 Roth's \cite{Roth} result that no strategyproof mechanism for marriage markets is stable. Firstly, I show that this impossibility extends to convex roommate problems and convex marriage markets. Secondly, I repeat
  Alcalde and Barbera's \cite{alcalde1994top} result that stability can be replaced with individual rationality and efficiency in Roth's \cite{Roth} result. Alcalde and Barbera \cite{alcalde1994top} complement their impossibility result with a condition on preferences under which Roth's \cite{Roth} impossibility result is overturned. If the domain of preferences satisfies `top dominance' then marriage markets permit stable and strategyproof mechanisms. Theorem \ref{theorem: MC satisfies the axioms} can then be interpreted as a more direct complement to Alcalde and Barbera's \cite{alcalde1994top} impossibility result: if all agents' preferences are convex then roommate problems (as well as marriage markets) permit strategyproof, individually rational and efficient mechanisms. Proposition \ref{proposition: Alcalde Barbera positive} in Appendix \ref{appendix: case of three} extends Alcalde and Barbera's \cite{alcalde1994top} to the case of three roommates. As Alcalde and Barbera's \cite{alcalde1994top} work pertains to marriage problems, they do not cover this case.

 Root and Ahn \cite{RootAhn} and Bade and Root \cite{BadeRoot} take a mechanism design approach to roommate problems. They show that sequential dictatorships are the only group-strategyproof and efficient mechanisms for roommate problems for the respective cases that roommates cannot or can stay single. In a sequential dictatorship a first agent gets to freely choose any partner. The mechanism then grants dictatorship rights to a next agent. This new dictator chooses his most preferred partner out of all agents who remain unmatched. As long as at least three agents remain unmatched the mechanism designates dictators. In the case that agents may stay single (Bade and Root \cite{BadeRoot}) a sequential dictatorship may terminate with a unanimity rule for the last two agents.

Bartholdi and Trick \cite{BartholdiTrick} and Richter and Rubinstein \cite{RiRu} study
roommate problems with single peaked preferences where agents may not stay single.
Bartholdi and Trick \cite{BartholdiTrick}  complement the assumption of single peakedness with narcissism in the sense that each agent prefers partners that are most similar to him or herself to show the existence of stable matchings.
Richter and Rubinstein \cite{RiRu}  illustrate their concept of unrestricted equilibrium with the case of roommate problems with single peaked preferences.  They show that such problems need not have any unrestricted equilibria  and that the sets of stable and unrestricted equilibrium allocations are not nested.
Example \ref{example: not single} provides an efficient, strategyproof and non-dictatorial mechanism for the case studied by Bartholdi and Trick \cite{BartholdiTrick},
 and Richter and Rubinstein \cite{RiRu}.

The assumption of single-peakedness has also been introduced into other matching models: Bade \cite{BadeSP} defined the Crawler, a strategyproof, individually rational and efficient mechanism for housing markets with single-peaked preferences that differs from Gale's top trading cycles. Similarlty to the mechanism defined in the present paper, the Crawler uses a sequence of smallest possible Pareto improvements to arrive at a matching,
 %This mechanism, the Crawler, and musical chairs share the feature that an algorithm gradually updates the agents matches.
  Hosseini and Tamura \cite{HosseiniTamura1} show that the application of respectively the Crawler or Gale's top trading cycles to endowments that are drawn uniform at random yield identical random matching mechanisms. Tamura \cite{Tamura1} provides two different characterizations of the Crawler. Tamura \cite{Tamura2} characterizes the full set of obviously strategyproof, efficient and individually rational mechanisms on the single-peaked domain. Just as in the Crawler most agents will gradually improve their currently assigned house in any such mechanism. But differently from the Crawler some agents may at some junctures of Tamura's \cite{Tamura2} mechanisms jump to a much preferred house.  Beynier et al. \cite{Beynier1} and  \cite{Beynier2} take a computer science lens to housing markets with  single-peaked preferences: they respectively study decentralized matching and computational complexity.
  Considering marriage markets with single peaked preferences, Mandal \cite{Mandal}  shows that stability and incentive compatibility do not clash on a subdomain of the single-peaked preferences. Echenique, Root and Sandomirsky \cite{EchRooSando} study the specific case of marriage markets with single-peaked preferences where each agent top ranks the closest partner. This - quite strong - restriction on preferences allows Echenique, Root and Sandomirsky \cite{EchRooSando} to relax other aspects of their model: their theory applies to a continuum of agent as well as to multi-partner matching.

\section{Definitions}

There is a set of agents $N=\{1,\dots, n\}$.
An agent $i$ can form a \textbf{pair} with a different agent $j$, denoted $(i,j)$ or stay \textbf{single}, denoted by - interchangeably - $(i)$ and $(i,i)$.
A \textbf{submatching} is a disjoint collection of pairs and singles.
A \textbf{matching} is a submatching that lists each agent in $N$.
  A submatching $\nu$ may also be represented as a function $\nu:N'\rightarrow N'$ for some $N'\subseteq N$ with $\nu\circ \nu$ the identity $id:N'\to N'$, so  that  $i$ is matched to $j$ if and only if $j$ is matched to $i$. The identity $id$ is the matching according to which each agent is single.
  The set of matchings is $\Sigma$.

Agents have strict preferences (linear orders) over $N$. For an arbitrary agent $i$'s preference $\succsim_i$,  $j\succ_i j'$ means that $i$ strictly prefers being matched with $j$ to being matched with $j'$. If $i\succ_i j$ then agent $i$ would rather be single than match with $j$.
Due to the assumption of linear rankings agent $i$ weakly prefers $j$ to $j'$  ($j\succsim_i j'$) if $j\succ_i j'$ or $j=j'$.
For any preference $\succsim_{i}$
define $\ptop(\succsim_i)$ as agent $i$'s most-preferred partner. The notation $\succsim_i: j,j',k$ means that agent $i$ respectively ranks agents $j$, $j'$ and $k$ in first, second and third place. Any agent's preference over matchings is derived from her preference over partners: Agent $i$ prefers matching $\mu$ to $\mu'$ if $\mu(i)\succsim_i\mu'(i)$.

 A preference $\succsim_i$ is single-peaked, if $j<j'\leq \ptop(\succsim_i)\Rightarrow j'\succ_i j $  and $\ptop(\succsim_i)\leq j'<j\Rightarrow j'\succ_i j$. Due to the assumption of linear orders, the notions of convexity and single peakedness coincide in the present model.
 The domains of all single-peaked preferences and of all preferences are denoted by $\hat{\Omega}\subseteq \Omega^{grand}$. A profile $\succsim\in \hat{\Omega}$ is a \textbf{convex roommate problem}.  A generic domain is denoted by $\Omega$.

A matching $\mu$ is \textbf{blocked} by a coalition $N'\subseteq N$ at $\succsim$ if there exists a submatching $\nu:N'\to N'$ such that $\nu(i)\succsim_i \mu(i)$ for all $i\in N'$ and $\nu(j)\succ_j\mu(j)$ for some $j\in N'$. A matching that is not blocked by any singletons at $\succsim$ is \textbf{individually rational}. A matching that is not blocked by the grand coalition $N$ is \textbf{efficient}. If two matchings $\mu'\neq \mu$ are such that $\mu'(i)\succsim_i\mu(i)$ then $\mu'$ \textbf{Pareto dominates} $\mu$.\footnote{The assumption of linear orders implies that $\mu'(j)\succ_j \mu(j)$ must hold for some $j$ if $\mu'$ Pareto dominates $\mu$.}
 A matching that is not blocked by any coalitions of size one or two (singletons or pairs) is \textbf{stable}. Any stable matching is by definition individually rational. To see that stable matchings must be efficient say $\mu$ is not efficient.  So  there exists an alternative matching $\mu':N\to N$ with $\mu'(i)\succsim_i \mu(i)$ for all $i$ and $\mu'(j)\succ_j \mu(j)$ for some $j$. If $\mu'(j)=j$ so that $j$ is single at $\mu'$, then $\mu$ is blocked by the singleton $(j)$. If $\mu'(j)=j'\neq j$ then the pair $(j,\mu'(j))$ blocks $\mu$. In either case $\mu$ is not stable.

A \textbf{mechanism} is a function $f:\Omega \rightarrow \Sigma$ that maps each profile of preferences to a matching.
Such a mechanism $f$ is \textbf{strategyproof} if there is no preference profile $\succsim$,  agent $i\in N$ and deviation $\succsim'_{i}$ such that $f(\succsim_{i}',\succsim_{-i})(i)\succ_{i} f(\succsim)(i)$. The mechanism $f$ is respectively efficient, individually rational and stable if it maps each $\succsim$ to a matching $f(\succsim)$ that is efficient, individually rational and stable at that profile.

A \textbf{marriage market} for a partition of $N$ into sets of men $M$ and women $W$
is a preference profile
$\succsim$ where the preference $\succsim_m$ of each man $m\in M$ ranks all women and himself: $W\cup \{m\}$, while the preference $\succsim_w$ of each woman $w\in W$
ranks all men and herself $M\cup \{w\}$. All preferences are linear orders and the set of preference profiles in the marriage market is $\Omega(M,W)$.
  A submatching for the marriage market is a submatching for the corresponding roommate problem that never matches two men or two women. The set of all matchings for the marriage market is denoted $\Sigma(M,W)\colon=\{\mu\in \Sigma\mid i\in W\Leftrightarrow \mu(i)\in M\}$.  Efficiency, stability,  individual rationality and strategyproofness in marriage markets are defined just as the corresponding notions for roommate problems.

\section{Direction and Gender}
Under the assumption of  individual rationality convex roommate problems can be interpreted  as  marriage markets where any agent's desire to move upward or downward defines their ``gender''.  Lemma \ref{lemma: embedding} in particular shows that a matching is individually rational in a convex roommate problem if and only if it is individually rational in the associated marriage market. The same correspondence extends to stability and individual rationality.

\begin{lemma}\label{lemma: embedding} Fix a convex roommate problem $\hat{\succsim}\in \hat{\Omega}$ and a $\mu\in \Sigma$.
Define a marriage market  $\succsim\in \Omega(M,W)$ for    $M\colon=\{i\in N\mid \ptop(\hat{\succsim}_i)\leq i\}$ and $W\colon=\{i\in N\mid \ptop(\hat{\succsim}_i)>i\}$ such that   $\succsim_w$ and $\succsim_m$  are for each $w\in W$ and $m\in M$  the restrictions  of $\hat{\succsim}_w$ and $\hat{\succsim}_m$ to respectively  $M\cup \{w\}$ and $W\cup \{m\}$.

\begin{itemize}
\item[a)] $\mu$ is individually rational at $\hat{\succsim}$ $\Rightarrow$  $\mu\in \Sigma(M,W)$.
\item[b)] $\mu$ individually rational at $\hat{\succsim}$ $\Leftrightarrow$ $\mu$ is individually rational at $\succsim$.
\item[c)] $\mu$ is stable at $\hat{\succsim}$ $\Leftrightarrow$  $\mu$ is stable at $\succsim$.
\item[d)] $\mu$ is individually rational and efficient at $\hat{\succsim}$ $\Leftrightarrow$

$\mu$ is individually rational and efficient at $\succsim$.
\end{itemize}

\end{lemma}

\begin{proof}
 Say $\mu$ is individually rational at $\hat{\succsim}$. Choose two  men $m<m'$. Since $m\in M$, $m\geq top(\hat{\succsim}_m)$. Since $\{\succsim\}_m$ is single peaked
 $m\succ_m m'$ and $(m,m')\notin \mu$. Mutatis mutandis we see that $(w,w')\notin \mu$ holds for any two different women.  So $\mu \in \Sigma(M,W)$ and a) holds.
 Since stability implies individual rationality and since parts b) and d) explicitly assume individual rationality, part a) implies that we may restrict attention to matchings $\mu\in \Sigma(M,W)$.

 By the definition of $\succsim$ and of marriage problems, $\mu(i)\succsim_i i$ holds if and only if $\mu(i)\hat{\succsim}_i i$ implying b). To see c) note that the definition of $\succsim$ implies that a pair $(m,w)$ blocks (the individually rational) $\mu$ at $\succsim$ if and only if it blocks $\mu$ at $\hat{\succsim}$.
To see d) note that, by the definition of $\succsim$,  a matching $\mu'\in \Sigma(M,W)$ Pareto dominates $\mu$ at $\succsim$ if and only if it does so at $\hat{\succsim}$. Now fix a $\mu'$ outside $\Sigma(M,W)$, so that $(i,j)\in \mu'$ holds for some same-gender pair $(i,j)$. By the arguments in the proof of claim a) at least one of the two partners, say $i$ prefers being single to $j$. Since $\mu$ is individually rational we then also have $\mu(i)\hat{\succsim}_i i\hat{\succ}_i \mu'(i)$ and $\mu'$ cannot Pareto dominate $\mu$ at $\hat{\succsim}$. In sum, we see that an individually rational $\mu$ is efficient at $\succsim$ if and only if it is efficient at $\hat{\succsim}$, proving part d)
 \end{proof}

Lemma \ref{lemma: embedding} can be used to transfer any number of positive results from marriage markets to convex roommate problems. Convex roommate problems in particular always have stable matchings, the set of stable matchings forms a distributive lattice, and deferred acceptance pins down the men and women optimal matchings. But to avoid introducing concepts such as deferred acceptance, a lattice, etc.,  I formally state only the first of these results:

\begin{corollary}\label{theorem: stability} (Chung \cite{Chung})
Any convex roommate problem $\hat{\succsim}\in \hat{\Omega}$ has a stable matching.
\end{corollary}

\begin{proof}
By Lemma \ref{lemma: embedding} the set of stable matchings at $\hat{\succsim}$ equals the set of stable matchings in the marriage market $\succsim$ derived from $\hat{\succsim}$ such that $M\colon=\{i\in N: \ptop(\hat{\succsim}_i)\leq i\}$, $W\colon=\{i\in N: \ptop(\hat{\succsim}_i)>i\}$, and such that $\succsim_m$ is for each $m\in M$ the restriction of $\hat{\succsim}_m$ to $W\cup \{m\}$ and $\succsim_w$ is for each $w\in W$ the restriction of $\hat{\succsim}_w$ to $M\cup \{w\}$. By Gale and Shapley \cite{GaleShapley} every marriage market has a stable matching.
\end{proof}

Chung \cite{Chung} shows Corollary \ref{theorem: stability} using a very different path: First, he shows that any roommate problem that has ``no odd rings'' has a stable matching. To demonstrate the usefulness of the ``no odd rings'' condition, Chung \cite{Chung} shows that convex roommate problems satisfy this condition.

Lemma \ref{lemma: embedding} does not mean that  convex roommate problems  and marriage markets are identical. On the one hand the sets of ``men'' and ``women''  arise endogenously in convex roommate problems. The assumption of single peaked preferences, on the other hand, carries over from any convex roommate problem into the associated marriage market. The latter difference has the - possibly surprising - implication that strategyproofness, efficiency and individual rationality can coexist in convex  roommate problems.
   To set the stage for this possibility result, Proposition \ref{proposition: stable} lists two variations of Roth's \cite{Roth} result on the incompatibility of strategyproofness and stability in marriage markets.

   Roth's \cite{Roth} result continues to hold if we either restrict attention to convex preferences or weaken stability to individual rationality and efficiency. The latter result is due to Alcalde and Barbera \cite{alcalde1994top}.

\begin{proposition}\label{proposition: stable}

\begin{itemize}
\item[a)] No stable mechanism for convex roommate problems with at least four agents  is strategyproof.
\item[b)] (Alcalde and Barbera \cite{alcalde1994top}) No individually rational and efficient mechanism for marriage markets with at least two men and two women is strategyproof.
\end{itemize}
\end{proposition}

\begin{proof}
Consider the case that there are  four agents $N=\{1,2,3,4\}$.

\noindent a) Fix a stable  $f:\hat{\Omega}\to \Sigma$, a profile $\succsim$ and two deviations $\succsim'_{2,4}$ such that
\begin{eqnarray*}
&\succsim_1: 4,3,2,1&\\
&\succsim_2: 3,4,2,1&~~~~\succsim'_2: 3,2,1,4\\
&\succsim_3: 1,2,3,4&\\
&\succsim_4: 2,1,3,4&~~~~ \succsim'_4: 2,3,4,1.
\end{eqnarray*}
Since $f$ is stable it  maps
  $\succsim$ to one of the two stable matchings at $\succsim$:  $\mu\colon=\{(1,4),(2,3)\}$ and $\mu'\colon=\{(1,3),(2,4)\}$. If $f(\succsim)=\mu$, agent 4 has an incentive to misrepresent her preferences as $\succsim'_4$: at $(\succsim'_4,\succsim_{-4})$ only $\mu'$ is stable,  and $f(\succsim'_4,\succsim_{-4})(4)=\mu'(4)=2\succ_4 1=\mu(4)$. Conversely if $f(\succsim)=\mu'$, agent $2$ has an incentive to misrepresent his preferences as $\succsim'_2$.

\noindent b) Fix a strategyproof, individually rational and efficient mechanism  $f:\Omega(M,W) \to \Sigma(M,W)$ for the case that $M=\{1,2\}$ and $W=\{3,4\}$. Define $\succsim^*$ as the restriction of $\succsim$ to $M$ and $W$:
\begin{eqnarray*}
&\succsim^*_1: 4,3,1\\
&\succsim^*_2: 3,4,2\\
&\succsim^*_3: 1,2,3\\
&\succsim^*_4: 2,1,4.
\end{eqnarray*}
The two matchings $\mu$ and $\mu'$, defined above, are the only individually rational and efficient matchings at $\succsim^*$.  The mechanism $f$  therefore maps $\succsim^*$  to $\mu$ or $\mu'$. Say $f(\succsim^*)= \mu$. Define   preferences $\succsim''_{3,4}$ for the women, so that they would rather be single than marry their
 partners according to $\mu$, keeping all else equal:
\begin{eqnarray*}
\succsim''_3: 1,3,2, ~~~~\succsim''_4: 2,4,1.
\end{eqnarray*}
To see that all agents must be single according to $f(\succsim''_{3,4},\succsim^*_{-3,4})$ we start by changing woman 4's preference from $\succsim^*_4$ to $\succsim''_4$.
\begin{eqnarray*}
&&\mbox{By strategyproofness }f(\succsim''_4,\succsim^*_{-4})(4)\in \{4,1\}.\\
&&\mbox{By individual rationality }f(\succsim''_4,\succsim^*_{-4})(4)\in \{4,2\}.\\
&&\mbox{So }f(\succsim''_4,\succsim^*_{-4})(4)=4.\\
&&\mbox{By feasibility }f(\succsim''_4,\succsim^*_{-4})\in \{id, \{(1),(2,3),(4)\},  \{(1,3),(2),(4)\}\}.\\
&&\mbox{By efficiency }f(\succsim''_4,\succsim^*_{-4})=\{(1),(2,3),(4)\}.
  \end{eqnarray*}
Changing woman 3's preference from $\succsim^*_3$ to $\succsim''_3$ yields the conclusion that 3 must be single in $f(\succsim''_{3,4},\succsim^*_{-3,4})$:
\begin{eqnarray*}
&&\mbox{By strategyproofness }f(\succsim''_{3,4},\succsim^*_{-3,4})(3)\in \{3,2\}.\\
&&\mbox{By individual rationality }f(\succsim''_{3,4},\succsim^*_{-3,4})(3)\in \{3,1\}.\\
&&\mbox{So }f(\succsim''_{3,4},\succsim^*_{-3,4})(3)=3.
  \end{eqnarray*}
     The application of the above arguments  to the sequence of profiles $\succsim^*$, $(\succsim''_3,\succsim^*_{-3})$ and $(\succsim''_{3,4},\succsim^*_{1,2})$  yields that  $f(\succsim''_{3,4},\succsim^*_{1,2})(4)=4$. In sum we then  see  $f(\succsim''_{3,4},\succsim^*_{1,2})=id$. A contradiction to efficiency, as $\mu'$ Pareto dominates  $id$ at $(\succsim''_{3,4},\succsim^*_{1,2})$. Mutatis mutandis the same arguments drive the alternative  assumption $f(\succsim^*)=\mu'$ to a contradiction.

To extend the two examples to any number of agents, define profiles $\succsim^{\circ}$ for $n>4$ such that $\ptop(\succsim_i)=i$ for all $i>4$ and such that the restrictions of $\succsim^{\circ}$ to $\{1,2,3,4\}$ coincide with the profiles defined above. In either case, the individual rationality of $f$  forces $f(\succsim^{\circ})(i)=i$ for any $i>4$ and the application of the above arguments to agents 1 through 4 then yields a contradiction.
\end{proof}

The impossibility results in parts a) and b) of Proposition \ref{proposition: stable} stem from a similar source: their proofs start with the same profile of preferences (for each agent $i$, $\succsim^*_i$ is the restriction of $\succsim_i$ to the agents of the other gender and him or herself). The reason why the claim in part b) does not transfer to convex roommate problems is that the proof of part b) relies on deviations that conflict with single-peakedness.   While the proof of Proposition \ref{proposition: stable} highlights the relation between the two impossibilities, Proposition \ref{proposition: 3 agents stable} in Appendix \ref{appendix: case of three} shows that  Alcalde and Barbera's \cite{alcalde1994top} result extends to roommate problems with just three agents.
Proposition \ref{proposition: stable} and \ref{proposition: 3 agents stable}
 seems to leave very little room for the existence of an individually rational, strategyproof and efficient mechanism for convex roommate problems. The next section constructs exactly such a mechanism.

\section{Dating}
Dating is an individually rational, strategyproof and efficient mechanism for convex roommate problems.
For any given profile of preferences all agents enter this mechanism as singles. The mechanism then gradually performs Pareto improvements by assigning agents to better and better dates. When multiple Pareto improvements are available at a step the mechanism always performs the ``smallest'' possible such improvement.
Whenever it becomes clear that some agent cannot be part of any further Pareto improvement such an agent is matched with their current date.

To define some further vocabulary
 fix a submatching $\nu:N'\to N'$ for a subset $N'\subset N$ together with a preference profile $\succsim$. An agent is stuck at $(\nu,\succsim)$ if there exists no other agent in $N'$ such that these two agents prefer each other to their roommates according to $\nu$, formally
agent $i\in N'$ is \textbf{stuck} at $(\nu,\succsim)$ if there exists no $j\in N'$ such that $i\succ_j \nu(i)$ and $j\succ_i \nu(i)$. So if agent $i$ is stuck at $(\nu,\succsim)$, then $N'$ either contains no agent $j$ whom $i$ prefers to $\nu(i)$, or any such agent $j$  prefers $\nu(j)$ to $i$. If $i$ is stuck, then there exists  no Pareto improvement $\nu'$ on $\nu$ with $\nu'(i)\neq \nu(i)$. An agent $i$ is \textbf{upwardly} (and respectively \textbf{downwardly}) mobile if there exists an agent $j>i$ ($j<i$) such that $j\succ_i \nu(i)$ and $i\succ_j \nu(j)$. By definition an agent cannot be stuck and either upwardly or downwardly mobile. If  $\succsim$ is a convex roommate problem then no agent is upwardly as well as downwardly mobile. So for any convex roommate problem $\succsim\in \hat{\Omega}$ and any submatching $\nu:N'\to N'$ we can partition the set $N'$ into the sets  $N_{up}$,  $N_{down}$ $N_{stuck}$  of upwardly mobile, downwardly mobile, and stuck agents.

A set $N^*\subseteq N'$ is a set of $N'$-\textbf{adjacent} agents if $N^*=\{i\in N'\mid \min N^*\leq i\leq \max N^*\}$.
Two agents $i,j\in N'$ and two sets disjoint sets $N^{\circ},N^*\subseteq N'$ are $N'$-adjacent if $\{i,j\}$ and $N^{\circ}\cup N^*$ respectively are sets of
 $N'$-adjacent agents.
A set $N^*\subseteq N'$ of $N'$-adjacent agents  is \textbf{framed} by two  agents $l,r\in N'$ with  $l<\min N^*$, $\max N^*<r$  if $\{l,r\}\cup N^*$ is a set of $N'$-adjacent agents.
  A \textbf{party} $P$ at a submatching $\nu:N'\to N'$ is a set of adjacent agents $P=L(P)\cup R(P)$ where  $R(P)\colon=\nu(L(P))$
  and  for all $i,j\in L(P)$ $i<\nu(i)$ and $i<j\Rightarrow \nu(i)<\nu(j)$. So $L(P)$ and $R(P)$ are the sets of left and right hand partners in the party $P$: if $i\in P$ and $i<\nu(i)$ then $i\in L(P)$ and $\nu(i)\in R(P)$. The set $\{1,2,3,4\}$ is, for example,  a party at the submatchings   $\{(1,2),(3,4)\}$ and $\{(1,3),(2,4)\}$ but not at   $\{(1,4),(2,3)\}$. To see the latter note that $1<2$, $1<\nu(1)$, $2<\nu(2)$ but $\nu(2)<\nu(4)$ holds for
    $\nu=\{(1,4),(2,3)\}$.
  Two adjacent parties  form a party.

\medskip

To define \textbf{dating} $D:\hat{\Omega}\to \Sigma$ fix any $\succsim\in \hat{\Omega}$ and use
the following algorithm to calculate $D(\succsim)$. Initialize the algorithm by setting $\nu^1:N^1\to N^1$ to  $id:N\to N$. As long as some agents remain unmatched, perform a step.

\bigskip

Step $k$.

\medskip

\textbf{Matching:} Match each agent $i$ who is stuck at $(\nu^k:N^k\to N^k,\succsim)$ with $\nu^k(i)$. If some agents are matched go to \textbf{Updating}, if not go to \textbf{Reassignment}

\medskip

\textbf{Reassignment:} For each party $P$ at $\nu^k$ that is framed by some $l\in N^k_{up}$ and $r\in N^k_{down}$ with $l<r$, define $\nu^{k+1}$ on $P\cup\{l,r\}$
 so that $P'$ with $L(P')=\{l\}\cup L(P)$ and $R(P')=\{r\}\cup R(P)$ is a party at $\nu^{k+1}$. Go to \textbf{Updating}.

\medskip

\textbf{Updating:} Update $N^{k+1}$ to the set of all agents that remain unmatched. Define $\nu^{k+1}:N^{k+1}\to N^{k+1}$ so that
 $\nu^{k+1}(i)\colon=\nu^k(i)$ for each agent $i\in N^{k+1}$ who was not reassigned at the current step.

\bigskip

Most of the above definition should be straightforward: Finding out whether  agent $i$ is  stuck at some step $k$ is easy: If agent $i$ top ranks $\nu^k(i)$ in the set $\{j\mid i\succsim_j \nu^k(j)\}$ of all agents $j$ who weakly prefer $i$ to their current dates, then $i$ is stuck.  Step $k$ then matches any such agent with their current date. If no agent is stuck
the algorithm moves to Reassignment.

Reassignment starts by screening for framed parties. For an example, say at some Step $k$ we had $N^k=\{1,2,5,6,11,13,14,15,17,20,21,30,31\}$. In Figure \ref{figure: parties} each of the agents is represented by a dot.
Say we also had $\nu^k(2)=2$, $\nu^{k}(5)=6$, $\nu^k(11)=14$, $\nu^k(13)=15$, and $\nu^k(17)=17$, so that at $\nu^k$ $\{5,6,11,13,14,15\}$ is party $P$ which is framed by the singles $2$ and $17$. The solid lines in  the upper half of   Figure \ref{figure: parties} connect $\nu^k$-dates  in the set $\{2,5,6,11,13,14,15,17\}$. For clarity the dates of the remaining agents are neither defined nor drawn. Assuming that agents 2 and 17 are respectively upwardly and downwardly mobile agents at $\nu^k$
  Step $k$  rearranges the partners of $\{2,17\}\cup P$ so that  $P'=\{2,5,6,11,13,14,15,17\}$ forms a party at $\nu^{k+1}$. The set $L(P')$  consists of agent 2 who frames the old party from below and (the old) $L(P)=\{5,11,13\}$. Analogously  $R(P')=\{6,14,15,17\}$. In sum, we get  $\nu^{k+1}(2)=6$, $\nu^{k+1}(5)=14$, $\nu^{k+1}(11)=15$ and $\nu^{k+1}(13)=17$. The solid lines in the lower half of  Figure \ref{figure: parties} connect all agents in the party $P'$ to their $\nu^{k+1}$-dates.

\begin{figure}[ht!]
\begin{tikzpicture}

\filldraw[black] (0,12) circle (1pt) node[label={[label distance=0.2cm]270:1}] {};
\filldraw[black] (1,12) circle (1pt) node[label={[label distance=0.2cm]270:2}] {};
\filldraw[black] (2,12) circle (1pt) node[label={[label distance=0.2cm]270:5}] {};
\filldraw[black] (3,12) circle (1pt) node[label={[label distance=0.2cm]270:6}] {};
\filldraw[black] (4,12) circle (1pt) node[label={[label distance=0.2cm]270:11}] {};
\filldraw[black] (5,12) circle (1pt) node[label={[label distance=0.2cm]270:13}] {};
\filldraw[black] (6,12) circle (1pt) node[label={[label distance=0.2cm]270:14}]{};
\filldraw[black] (7,12) circle (1pt) node[label={[label distance=0.2cm]270:15}] {};
\filldraw[black] (8,12) circle (1pt) node[label={[label distance=0.2cm]270:17}] {};
\filldraw[black] (9,12) circle (1pt) node[label={[label distance=0.2cm]270:20}]{};
\filldraw[black] (10,12) circle (1pt) node[label={[label distance=0.2cm]270:21}] {};
\filldraw[black] (11,12) circle (1pt) node[label={[label distance=0.2cm]270:30}] {};
\filldraw[black] (12,12) circle (1pt)  node[label={[label distance=0.2cm]270:31}] {};

%\draw (1.5,13) rectangle (2.5,11);

\draw[thick] (.95,12.01) ..controls (.7,12.5) and (1.3,12.5) .. (1.05,12.01);

\draw[thick] (2.05,12.01).. controls (2.2, 12.8) and (2.8,12.8).. (2.95,12.01);

\draw[thick] (4.05,12.01).. controls (4.3, 12.8) and (5.7,12.8).. (5.95,12.01);

\draw[thick] (5.05,12.01).. controls (5.3, 12.8) and (6.7,12.8).. (6.95,12.01);

\draw[thick] (7.95,12.01) ..controls (7.7,12.5) and (8.3,12.5) .. (8.05,12.01);

\filldraw[black] (0,10) circle (1pt) node[label={[label distance=0.2cm]270:1}] {};
\filldraw[black] (1,10) circle (1pt) node[label={[label distance=0.2cm]270:2}] {};
\filldraw[black] (2,10) circle (1pt) node[label={[label distance=0.2cm]270:5}] {};
\filldraw[black] (3,10) circle (1pt) node[label={[label distance=0.2cm]270:6}] {};
\filldraw[black] (4,10) circle (1pt) node[label={[label distance=0.2cm]270:11}] {};
\filldraw[black] (5,10) circle (1pt) node[label={[label distance=0.2cm]270:13}] {};
\filldraw[black] (6,10) circle (1pt) node[label={[label distance=0.2cm]270:14}]{};
\filldraw[black] (7,10) circle (1pt) node[label={[label distance=0.2cm]270:15}] {};
\filldraw[black] (8,10) circle (1pt) node[label={[label distance=0.2cm]270:17}] {};
\filldraw[black] (9,10) circle (1pt) node[label={[label distance=0.2cm]270:20}]{};
\filldraw[black] (10,10) circle (1pt) node[label={[label distance=0.2cm]270:21}] {};
\filldraw[black] (11,10) circle (1pt) node[label={[label distance=0.2cm]270:30}] {};
\filldraw[black] (12,10) circle (1pt)  node[label={[label distance=0.2cm]270:31}] {};

\draw[thick] (1.05,10.01) ..controls (1.3,10.8) and (2.7,10.8) .. (2.95,10.01);

\draw[thick] (2.05,10.01).. controls (2.5, 10.9) and (5.5,10.9).. (5.95,10.01);

\draw[thick] (4.05,10.01).. controls (4.4, 10.9) and (6.6,10.9).. (6.95,10.01);

\draw[thick] (5.05,10.01).. controls (5.4, 10.9) and (7.6,10.9).. (7.95,10.01);

\end{tikzpicture}

\caption{Reassignment from $\nu^k$ to $\nu^{k+1}$ }

\end{figure}
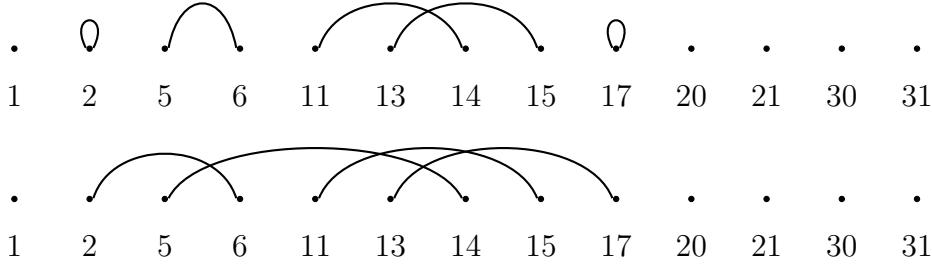\label{figure: parties}

Once matches and reassignments have been made, the algorithm moves to Updating. The instructions here are straightforward: any unmatched agents whose partners have not been updated continue to date the same people.

The present illustration made three assumptions for the case that no agent is stuck.  Firstly, I assumed that there exists a party. Secondly, I assumed that this party is framed by an upwardly mobile single from below and a downwardly mobile single from above. Third, I assumed that the instructions to rearrange the partners of the agents in the framed party at $\nu^k$ would produce a well-defined new party at $\nu^{k+1}$.  The proof that dating is well-defined revolves around showing that these assumptions hold true for any $\nu^k$  that can be reached via the algorithm.
One special case of reassignments is easy: Any two adjacent singles frame the ``empty party''. If no agent is stuck ($N_{stuck}^k=\emptyset$) at Step $k$, then Step $k$ reassigns any two such singles $j$ and $j'$
 to each other ($\nu^{k+1}(j')=j$) if  $j\in N^k_{up}$ and $j<j'\in N^k _{down}$. For a detailed example of the dating algorithm see Appendix \ref{appendix: mc example}.

\begin{theorem}\label{theorem: MC satisfies the axioms}
Dating $D:\hat{\Omega} \to \Sigma$ is welldefined, strategyproof, individually rational and efficient.
\end{theorem}

The proof of Theorem \ref{theorem: MC satisfies the axioms} crucially relies on
 Lemma \ref{the lemma}  describing the steps of  the dating algorithm. Call the algorithm used to calculate the outcome of $D$ at  $\succsim$, the $D(\succsim)$-algorithm. Lemma \ref{the lemma} shows that it is always possible to perform a step, that agents never get reassigned to worse partners, and that reassignments are unidirectional. The proof of Lemma \ref{the lemma} is in Appendix \ref{appendix: Lemma proof}.

\begin{lemma}\label{the lemma}
Fix a profile $\succsim\in \hat{\Omega}$.
Each Step $k$ of the $D(\succsim)$-algorithm either matches or reassigns some agents. Each current submatching $\nu^{k}$ is made up of singles and parties, and  $L(P)\cap N^k_{down} =\emptyset= R(P)\cap N^k_{up}$ holds for any party $P$ at $\nu^k$.  Any $i\in N^{k+1}$ who is reassigned to a new partner $\nu^{k+1}(i)\neq \nu^k(i)$ strictly prefers that new partner $ \nu^{k+1}(i)\succ_i \nu^k(i)$ and  $(\ptop(\succsim_i)-i)( \nu^{k+1}(i)- \nu^k(i))>0$ .
\end{lemma}

\begin{proof}\textbf{of Theorem \ref{theorem: MC satisfies the axioms}}
Fix an arbitrary $\succsim\in \hat{\Omega}$.

\medskip

 $D$ is \textbf{well-defined}. By Lemma \ref{the lemma}, each step of the $D(\succsim)$-algorithm that does not math any agent must
 reassign some agents to strictly preferred partners.  Since there are only finitely many agents, there are only finitely many possible improvements. Once all agents are matched, the mechanism terminates.

\medskip

$D$ is \textbf{individually rational}. Fix an agent $i$ and say this agent gets matched at Step $k^*$. If $k^*=1$, then agent $i$ gets matched with $\nu^1(i)=i$. If $k^*>1$ then
 $ \nu^{k+1}(i)\succsim_i \nu^{k}(i)$ holds (by Lemma \ref{the lemma}) for any Step  $k<k^*$ of the $D(\succsim)$-algorithm. Transitivity together with
 $\nu^1(i)=i$  then imply $D(\succsim)(i)=\nu^{k^*}\succsim_i i=\nu^1(i)$ and $D$ is individually rational.

\medskip

$D$ is \textbf{efficient}. Fix a matching $\mu\neq D(\succsim)$. Say that Step $k^*$ is the first step where $D(\succsim)$ finds a match $(i,i')\notin \mu$.
Since the $D(\succsim)$-algorithm matches $i$ with $i'$ at Step $k^*$, at least one of these two agents, say $i$, must be stuck at ($\nu^{k^*},\succsim)$.  Say   $(i,i'')\in \mu$.
 %so that
 %$D(\succsim)(i)=i'=\nu^{k^*}(i)\succsim_i j$ for all $j\in N^{k^*}$ with $i\succsim_j \nu^{k^*}(j)$.
 %Say   $(i,i'')\in \mu$.
 Since $i$ is stuck at  $(\nu^{k^*},\succsim)$, since $\nu^{k^*}(i)=i'$ and since
 $i''\in N^{k^*}$ we have  either a) $i'\succ_i i''$ or b) $\nu^{k^*}(i'')\succ_{i''} i$.
 In either case $\mu$ cannot Pareto dominate $D(\succsim)$  as a) yields $D(\succsim)(i)=i'\succ_i \mu(i)=i''$ and b) in combination with Lemma \ref{the lemma} yields  $D(\succsim)(i'')\succsim_{i''}\nu^{k^*}(i'')\succ_{i''}\mu(i'')=i$. (Note that the proof did not rule our $i=i'$ or $i=i''$.)

\medskip

$D$ is \textbf{strategyproof}.
Suppose there exist an agent $i$, a profile $\succsim$, and a deviation $\succsim'_i$ such that $ D(\succsim'_i,\succsim_{-i})(i)\succ_i D(\succsim)(i)$.
Since $D$ is individually rational, $\ptop(\succsim_i)=$ implies $D(\succsim)(i)=top(\succsim_i)$. We may therefore w.l.o.g assume that $\ptop(\succsim_i)>i$, noting that all upcoming arguments apply mutatis mutandis to the case that $\ptop(\succsim_i)<i$.
By individual rationality and single peakedness we have that $D(\succsim''_i,\succsim_i)(i)\leq i\leq D(\succsim)(i)$ for any $\succsim''_i$ with $top(\succsim''_i)<i$. So  $ D(\succsim'_i,\succsim_{-i})(i)\succ_i D(\succsim)(i)$ can only hold if $top(\succsim'_i)>i$.

  Say $k^*$ is the first step where the $D(\succsim)$- and $D(\succsim'_i,\succsim_{-i})$-algorithms differ. Since the two algorithms proceed identically up to Step $k^*$, Step $k^*$ of both algorithms starts with the same current matching $\nu^{k^*}:N^{k^*}\to N^{k^*}$. If $\nu^{k^*}(i)$ is stuck at Step $k^*$ then  we obtain the contradiction that the $D(\succsim)$- and $  D(\succsim'_i,\succsim_{-i})$-algorithms both match $i$ at Step $k^*$ with $\nu^{k^*}(i)$.

Case 1:
 The $D(\succsim'_i,\succsim_{-i})$-algorithm  matches agent $i$ at Step $k^*$.  So $D(\succsim'_i,\succsim_{-i})(i)=\nu^{k^*}(i)$. By Lemma \ref{the lemma} the $D(\succsim)$-algorithm reassigns agent $i$ to weakly preferred partners and ultimately matches $i$ with such a weakly preferred partner $D(\succsim)(i)$ - a contradiction to $ D(\succsim'_i,\succsim_{-i})(i)\succ_i D(\succsim)(i)$. Case 2: $D(\succsim)$-algorithm matches agent $i$ at Step $k^*$. Since agent $i$ is stuck at $(\nu^{k^*},\succsim)$: $\nu^{k^*}(i)\succsim_i j$ for all $j\in N^{k^*}$ with $i\succsim_j \nu^{k^*}(j)$. By Lemma \ref{the lemma} $D(\succsim'_i,\succsim_{-i})$ can only match
 agent $i$ with  an agent $j$ satisfying the latter condition - a contradiction to $ D(\succsim'_i,\succsim_{-i})(i)\succ_i D(\succsim)(i)$.

 Given that Cases 1 and 2 do not hold, neither of the two algorithms matches agent $i$ at Step $k^*$. For the two algorithms to differ exactly one of the two must reassign agent $i$ at Step $k^*$. So at  $\nu^{k^*}$, there exists a party $P$ framed by two singles $l<r$ such that at $\succsim$: $i\in \{l\}\cup L(P)$, $(\{l\}\cup L(P))\setminus \{i\}\subset  N_{up}$ and $\{r\}\cup R(P)\subset N_{down}$. Say the algorithm which reassigns agent $i$ to a new partner at Step $k^*$ reassigns her to $\nu^{k^*+1}(i)\neq \nu^{k^*}(i)$. For the other algorithm not to reassign agent $i$ at Step $k^*$, agent $i$ must, in that other algorithm strictly prefer $\nu^{k^*}(i)$ to $\nu^{k^*+1}(i)$. But that means that agent $i$ is according to the other algorithm stuck at Step $k^*$ - a contradiction to Cases 1 and 2 above which showed that neither algorithm matches agent $i$ at Step $k^*$.

\end{proof}

\section{Different mechanisms, Different domains}
\subsection{Convex Marriage Markets}
Alcalde and Barbera \cite{alcalde1994top} showed that stability in  Roth's \cite{Roth} impossibility result for marriage markets can be replaced with incentive compatibility and efficiency. They complemented this non-existence result with a domain restriction for which the exist stable and strategyproof mechanisms for marriage markets. Proposition \ref{proposition: Alcalde Barbera positive} is a more direct complement to Alcalde and Barbera's \cite{alcalde1994top} non-existence result.

\begin{proposition}\label{proposition: Alcalde Barbera positive}
There are efficient, individually rational and strategyproof mechanisms for marriage markets with single peaked preferences.
\end{proposition}

The proof of proposition adapts dating to marriage markets by splitting the set of agents into two separate dating pools.

\begin{proof}
To define an efficient, individually rational and strategyproof mechanism for marriage markets with single peaked preferences adapt dating $D$ to $D'$ as follows. For any marriage market $\succsim$ partition $W$ and $M$ into agents who prefer partners above and below themselves, so
\begin{eqnarray*}
M^u\colon=\{i\in M\mid \ptop(\succsim_i)\geq i\}, W^u\colon=\{i\in W\mid \ptop(\succsim_i)\geq i\}\\
M^d\colon=\{i\in M\mid \ptop(\succsim_i)< i\}, W^d\colon=\{i\in W\mid \ptop(\succsim_i)< i\}.
\end{eqnarray*}
Define two new problems. The first comprises the men $M^u$ and the women $W^d$, the second the men $M^d$ and the women $W^u$. Say $\succsim^1$ and $\succsim^2$ respectively are the restrictions of $\succsim$ to $N^1\colon=M^u\cup W^d$ and $N^2\colon=M^d\cup W^u$. Define $D'(\succsim)$ as the union of $D(\succsim^1)$ and $D(\succsim^2)$.
\end{proof}

Alcalde and Barbera's \cite{alcalde1994top} of top dominance  and single peakedness do not imply each other: While top dominance guarantees the existence of stable and strategyproof mechanisms for roommate problems part a) of Proposition \ref{proposition: stable} shows that single peakedness is not sufficient for the existence of such mechanisms. On the other hand, Alcalde and Barbera \cite{alcalde1994top} give an example of preferences that are not single peaked but satisfy top dominance.  Since stability implies individual rationality and efficiency it is then clear that the domain of preferences for which there exist strategyproof, individually rational and efficient  mechanisms for marriage markets is larger than the domain of single peaked preferences.
\subsection{Other mechanisms for convex roommate problems}

Dating $D$ is not the only individually rational, strategyproof and efficient mechanism for convex roommate problems: Example \ref{example: more mechanisms} constructs an alternative mechanism $\tilde{D}$ with these properties. Roughly speaking $\tilde{D}$ and  $D$ differ insofar as they  prioritize the formation of pairs among  either the most distant or the closest agents. Where $D$ incrementally improves the lot of as many agents as possible, $\tilde{D}$ yields more substantial improvements to fewer agents.  For any $\succsim$ the largest distance between partners in $\tilde{D}(\succsim)$ is at least as large as the largest such distance among partners in $D(\succsim)$. For simplicity I only define $\tilde{D}$ for sets of three agents.

\begin{example}\label{example: more mechanisms}
Fix $N=\{1,2,3\}$. To define  $\tilde{D}:\hat{\Omega}\to \Sigma$ fix an arbitrary $\succsim\in \hat{\Omega}$. Say $\overline{\Sigma}(\succsim)$ is the set of all individually rational and efficient matchings at $\succsim$. If $\overline{\Sigma}(\succsim)$ is a singleton $\{\mu\}$, set $\tilde{D}(\succsim)$ to $\mu$. If not choose $\mu$ such that $\mu(i)=\ptop(\succsim_i)$ for $i\succ_2 2$.
To see that $\tilde{D}$ is welldefined note that $\ptop(\succsim_2)\neq 2$ must hold for
$\overline{\Sigma}(\succsim)$  not to be  a singleton. If $\overline{\Sigma}(\succsim)$ is   a singleton, the efficiency and individual rationality of $\tilde{D}(\succsim)$ follow directly from its definition. If not, say $i\succ_2 2$. Since $\tilde{D}(\succsim)(i)=\ptop(\succsim_i)\neq i$, $\tilde{D}(\succsim)$ is then efficient. Since, in addition, $i\succ_j i$ holds for $j\neq i$, $\tilde{D}(\succsim)$ is individually rational.

The proof that $\tilde{D}$ is strategyproof is divided in two Cases. Case 1:   $\ptop(\succsim_i)=i$ for some $i\in \{1,2,3\}$. For this agent $\tilde{D}(\succsim)(i)=i$ and the two agents $\{1,2,3\}\setminus \{i\}$ are paired if and only if they prefer this pairing to staying single. So no agent has an incentive to deviate in Case 1. Case 2:
$\ptop(\succsim_i)\neq i$ for $i\in\{1,2,3\}$.  Say $\ptop(\succsim_2)=1$. Since $\tilde{D}$ is individually rational $\tilde{D}(\succsim)(2)\in \{1,2\}$ while $\tilde{D}(\succsim'_2,\succsim_{-2})(2)\in \{2,3\}$ for $\succsim'_2\neq \succsim_2$. The individual rationality of $\tilde{D}$ also implies that $\tilde{D}(\succsim)(3)\in \{1,3\}$ and that $\tilde{D}(\succsim)(3)$ can only equal $1$ if $1\succ_3 3$. So agents 2 and 3 do not have any incentive to deviate.
Agent 1 finally does not have an incentive to deviate since $\tilde{D}(\succsim)$ matches agent 1 with agent 1's top ranked partner among all agents $i$ with $1\succ_i i$.  Mutatis mutandis the same arguments apply when $\ptop(\succsim_2)=3$.

Dating $D$ and the new mechanism $\tilde{D}$ differ as  $\tilde{D}(\succsim^*)=\{(1,3),(2)\}$ and $D(\succsim^*)=\{(1), (2,3)\}$ holds for  $\succsim^*$ with $\ptop(\succsim^*_1)=\ptop(\succsim^*_2)=3$ and $\ptop(\succsim^*_3)=1$.
\end{example}

\subsection{The case where agents may not stay single}
 Example \ref{example: not single} applies to roommate problems where all agents must be paired as studied by Root and Ahn \cite{RootAhn} and Rubinstein and Richter \cite{RiRu}. Also in this case the restriction to single peaked preferences increases the set of efficient and
strategyproof mechanisms.
 Say that $\Omega^{pair}$ and $\hat{\Omega}^{pair}$ respectively are the grand and the single peaked domains of all linear preference profiles for roommate problems where agents must be paired.

\begin{example}\label{example: not single}
Say $N=\{1,2,3,4\}$. The three possible matchings $\mu^1=\{(1,2),(3,4)\}$, $\mu^2=\{(1,3),(2,4)\}$ and $\mu^3=\{(1,4),(2,3)\}$ are illustrated in Figure \ref{figure: m-pair}.
Define a mechanism $D^{pair}:\hat{\Omega}^{pair} \to \{\mu^1,\mu^2,\mu^3\}$ so that a) $D^{pair}(\succsim)= \mu^3$ if $\mu^3$ is efficient at $\succsim$ and b) $D^{pair}(\succsim)=\mu^k$ if $\mu^k$ Pareto dominates $\mu^3$ at $\succsim$.
The mechanism $D^{pair}$ is well-defined since there is no $\succsim$ at which $\mu^1$ and $\mu^2$ both Pareto dominate $\mu^3$: by single-peakedness
we either have $\mu^1(2)=1\succ_2 \mu^3(2)=3\succ_ 2 \mu^2(2)=4$ or $4\succ_2 3\succ_ 2 1$.

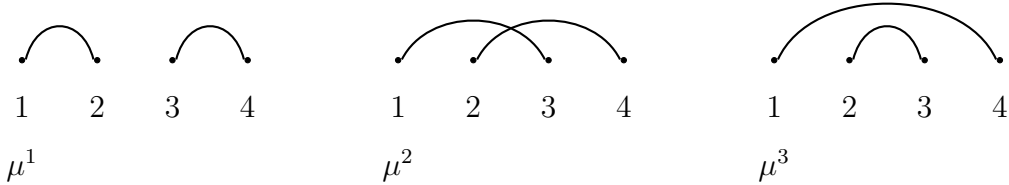
\begin{figure}[ht!]
\begin{tikzpicture}

\filldraw[black] (0,1) circle (1pt) node[label={[label distance=0.2cm]270:1}] [label={[label distance=.9cm]270:$\mu^1$}]{};
\filldraw[black] (1,1) circle (1pt) node[label={[label distance=0.2cm]270:2}] {};
\filldraw[black] (2,1) circle (1pt) node[label={[label distance=0.2cm]270:3}] {};
\filldraw[black] (3,1) circle (1pt) node[label={[label distance=0.2cm]270:4}] {};
%\filldraw[black] (4,1) circle (1pt) node[label={[label distance=0.2cm]270:11}] {};
\filldraw[black] (5,1) circle (1pt) node[label={[label distance=0.2cm]270:1}] [label={[label distance=.9cm]270:$\mu^2$}] {};
\filldraw[black] (6,1) circle (1pt) node[label={[label distance=0.2cm]270:2}]{};
\filldraw[black] (7,1) circle (1pt) node[label={[label distance=0.2cm]270:3}] {};
\filldraw[black] (8,1) circle (1pt) node[label={[label distance=0.2cm]270:4}] {};
%\filldraw[black] (9,1) circle (1pt) node[label={[label distance=0.2cm]270:20}]{};
\filldraw[black] (10,1) circle (1pt) node[label={[label distance=0.2cm]270:1}] [label={[label distance=.9cm]270:$\mu^3$}] {};
\filldraw[black] (11,1) circle (1pt) node[label={[label distance=0.2cm]270:2}] {};
\filldraw[black] (12,1) circle (1pt)  node[label={[label distance=0.2cm]270:3}] {};
\filldraw[black] (13,1) circle (1pt)  node[label={[label distance=0.2cm]270:4}] {};

%\draw[thick] (.95,12.01) ..controls (.7,12.5) and (1.3,12.5) .. (1.05,12.01);

\draw[thick] (0.05,1.01).. controls (0.2, 1.6) and (0.8,1.6).. (0.95,1.01);

\draw[thick] (2.05,1.01).. controls (2.2, 1.6) and (2.8,1.6).. (2.95,1.01);

\draw[thick] (5.05,1.01).. controls (5.4, 1.7) and (6.6,1.7).. (6.95,1.01);

\draw[thick] (6.05,1.01).. controls (6.4, 1.7) and (7.6,1.7).. (7.95,1.01);

\draw[thick] (10.05,1.01).. controls (10.5, 2) and (12.5,2).. (12.95,1.01);

\draw[thick] (11.05,1.01).. controls (11.2, 1.6) and (11.8,1.6).. (11.95,1.01);

\end{tikzpicture}
\caption{The three possible matchings}
\end{figure}\label{figure: m-pair}

The mechanism $D^{pair}$ is efficient by definition.
To see that  $D^{pair}$ is strategyproof fix $\succsim,\succsim'_i$ and $\succsim''_i$ such that $D^{pair}(\succsim'_i,\succsim_{-i})(i)\neq D^{pair}(\succsim''_i,\succsim_{-i})(i)$. If  $D^{pair}(\succsim'_i,\succsim_{-i})=\mu^1$ and $D^{pair}(\succsim''_i,\succsim_{-i})=\mu^2$, then $\mu^1$ and $\mu^2$ must Pareto dominate $\mu^3$ at respectively at $(\succsim'_i,\succsim_{-i})$ and $(\succsim''_i,\succsim_{-i})$. Since $D^{pair}$ treats agents 2 and 3 symmetrically, we may w.l.o.g assume that $i\neq 2$. For the Pareto dominance relations to hold we then need $D(\succsim'_i,\succsim_{-i})(2)=\mu^1(2)=1\succ_2 \mu^3(2)=3$ as well as $D(\succsim''_i,\succsim_{-i})(2)=\mu^2(2)=4\succ_2 \mu^3(2)=3$ - a contradiction to $\succsim_2$ being single peaked.
So  $D^{pair}(\succsim'_i,\succsim_{-i})(i)\neq D^{pair}(\succsim''_i,\succsim_{-i})(i)$ implies that one of these two matchings equals $\mu^3$.

So say that $D^{pair}(\succsim'_i,\succsim_{-i})=\mu^3$ and $ D^{pair}(\succsim''_i,\succsim_{-i})=\mu^{k}$ for $k\in\{1,2\}$.  By the definition of $D^{pair}$ $\mu^{k}$ must Pareto dominate $\mu^3$ at $(\succsim''_i,\succsim_{-i})$. So $f(\succsim''_i,\succsim_{-i})(i)=\mu^{k}(i)\succ''_i\mu^3(i)=f(\succsim'_i,\succsim_{-i})(i)$.
Conversely $\mu^{k}$ cannot Pareto dominate $\mu^3$ at $(\succsim'_i,\succsim_{-i})$. Since agent $i$'s preference is the only difference between the profiles $(\succsim'_i,\succsim_{-i})$ and $(\succsim''_i,\succsim_{-i})$, we see that $\mu^3(i)=f(\succsim'_i,\succsim_{-i})(i)\succ'_i\mu^k(i)=f(\succsim''_i,\succsim_{-i})(i)$ and $\mu^k(i)=f(\succsim''_i,\succsim_{-i})(i)\succ''_i\mu^3(i)=f(\succsim'_i,\succsim_{-i})(i)$ and $D^{pair}$ is strategyproof.

Finally, the grand domain $\Omega^{pair}$ lets the four agents express any (strict) ranking over the three possible matchings $\{\mu^1,\mu^2,\mu^3\}$.  By the Gibbard \cite{Gibbard} and Satterthwaite \cite{Satterthwaite} theorem, the only efficient and strategyproof mechanism for this problem is a dictatorship. But $D^{pair}$ cannot be a dictatorship since it chooses $\mu^3$ whenever at least one agent $i$ top ranks $\mu^3(i)$.
\end{example}

On the domain of all linear orders we can derive strategyproof and efficient mechanisms for the case that agents cannot stay single from such mechanisms for the case that agents can stay single. In that case the restriction $\overline{f}$ of any strategyproof and efficient mechanism $f:\Omega^{grand}\to \Sigma$ to the domain $\overline{\Omega}$  where all agents rank being single at the bottom defines a strategyproof and efficient mechanism for the case where agents cannot be single. Since $f$ is efficient, the restriction $\overline{f}$ of $f$ to $\overline{\Omega}$ never leaves any agent single. Since $f$ is efficient and strategyproof, $\overline{f}$ is so too.
 The reason why this embedding strategy does not work on the domain of single peaked preferences $\hat{\Omega}$ is that this domain
 does not include a subdomain where all agents rank being single at the bottom. For an example say $N=\{1,2,3,4\}$ and $\succsim_3\in \hat{\Omega}_3$ top ranks agent 4. Then $\hat{\Omega}_3$ does not contain a preference $\succsim'_1$ that top ranks $4$ and bottom ranks $3$, as no such preference is single peaked.

\section{Conclusion}

Within roommate problems the assumption of convex preferences opens up many doors. While strategyproofness, efficiency and individual rationality are incompatible in standard roommate problems,  dating satisfies all three  when agents have single-peaked preferences (Theorem \ref{theorem: MC satisfies the axioms}). While generic roommate problems may lack stable matchings, roommate problems with single peaked preferences always permit stable matchings (Corollary \ref{theorem: stability}).

Examples \ref{example: more mechanisms} and \ref{example: not single} raise a set of questions about further possibilities for convex roommate problems. Since Example \ref{example: more mechanisms} presents an alternative strategyproof, efficient and individually rational mechanism for convex roommate problems, one may wonder about the full set of mechanisms with these three properties. A standard trick can be used to leverage the dating mechanism $D$ and the version $\tilde{D}$, defined in Example \ref{example: more mechanisms},  to construct many more mechanisms: Consider the case of $N=\{1,2,3,4\}$ and always use $D$ to match all agents, except when agent 2 has the preference $\succsim^*_2: 2,1,3,4$.  In that case let agent 2 be single and use $\tilde{D}$ to match agents 1,3 and 4.
  The resulting mechanism can be manipulated by groups: fix a profile $\succsim$ with $\succsim_2=\succsim^*_2$ and assume that agent 1 prefers that $D$ be used to match agents $\{1,3,4\}$.
   Agents 1 and 2 then have a joint deviation $\succsim'_{1,2}$ which keeps agent 2 single (his optimum) and improves agents 1's match: Agent 2 should announce $\succsim'_1: 2,3,4,1$ and agent 1 should keep their preference $\succsim'_1=\succsim_1$.
    Mechanisms generated via this cheap trick are  generally not ``group-strategyproof''.

But some preliminary evidence suggests that even the set of group-strategyproof, efficient and individually rational  mechanisms  is large. This set may even contain a large subset of
mechanisms that are  ``obviously strategyproof'' according to the definition by Li \cite{Li}. The intuition for this claim comes from Tamura's \cite{Tamura2}'s characterization of all obviously strategyproof and efficient mechanisms for housing markets with single-peaked preferences. A mechanism should be obviously strategy proof if it  uses an algorithm where each agent who gets to make a choice either immediately leaves the mechanism or chooses to update her safe option to a minimally improved safe option.

The option of staying single plays a crucial role in the dating mechanism. Example \ref{example: not single}'s mechanism $D^{pair}$ for the case where agents may not stay single differs indeed markedly from dating $D$. In a sense we could view $D^{pair}$ as the polar opposite of $D$. Each agent enters the dating mechanism
with their closest possible partner (him or herself). Conversely the fallback matching $\mu^3$ starts with the most distant pair possible: $(1,4)$. Only if all agents find a deviation from this matching beneficial will $D^{pair}$ not match  $1$ and $4$. The question whether all convex roommate problems with the option to stay single permit such a mechanism is open.
 On a more grand level there is the question how to relate the sets of mechanisms with certain properties for roommate problems with and without the option to stay single. For the two domains of all linear orders, Root and Ahn \cite{RootAhn} and Bade and Root \cite{BadeRoot} use remarkably different proofs to find nearly identical characterizations of all group-strategyproof and efficient mechanisms for roommate problems. It would be nice to find strategies of proof that can simultaneously cover these two arguably so similar problems.

\appendix

\section{A 3-agent impossibility result}\label{appendix: case of three}

The following proposition shows that there exists no  strategyproof, individually rational and efficient mechanism for roommate problems with three agents. The proof amends
 Gale and Shapley's \cite{GaleShapley}   famed example of a roommate problem without a stable matching.

\begin{proposition}\label{proposition: 3 agents stable}
There exists no strategyproof, individually rational and efficient mechanism $f:\Omega^{grand} \to \Sigma$ for $N=\{1,2,3\}$.
\end{proposition}

\begin{proof}
Suppose $f:\Omega^{grand} \to \Sigma$ was   strategyproof and individually rational. I first show that any agent $i$ who is top ranked by some different agent $j$ will be able to get a match he or she weakly prefers to $j$.
  Say $\ptop(\succsim_j)=i$, $\succsim'_i: j,i$ and $\succsim'_j: i,j$. Since $f$ is individually rational either $(i),(j)\in f(\succsim'_{i,j},\succsim_{-i,j})$ or $(i,j)\in f(\succsim'_{i,j},\succsim_{-i,j})$. Since $i$ and $j$ both prefer being matched with each other to being single  $(i,j)\in f(\succsim'_{i,j},\succsim_{-i,j})$ holds by efficiency. By strategyproofness $(i,j)\in f(\succsim'_{i},\succsim_{-i})$. Applying strategyproofness once more we see that
$ f(\succsim)(i)\succsim_i j$.

So $f(\succsim)(1)\succsim_3$, $f(\succsim)(2)\succsim_2 1$, and $f(\succsim)(3)\succsim_3 2$ holds for $\succsim_1: 2,3,1$, $\succsim_2: 3,1,2$, $\succsim_3: 1,2,3$.
The latter implies that each $i$ strictly prefers $f(\succsim)(i)$ to being single.
 A contradiction arises since  $f(\succsim)(i)=i$ must hold for at least one agent $i$.
\end{proof}

\section{An example of the dating algorithm}\label{appendix: mc example}

Say there are six agents $\{1,2,\dots, 6\}$. Assume throughout that  agents 1 and 2 always prefer higher partners  whereas agents 5 and 6 prefer lower partners ($\succsim_i: 6,5,4,3,2,1$ for $i=1,2$ and $\succsim_i: 1,2,3,4,5,6$ for $i=5,6$).
Since agents 1 and 2 prefer agents 5 and 6 to their current partner (themselves) and vice versa, agents 1, 2, 5, and 6 are not stuck (and therefore not matched) at Step 1. This conclusion does not depend on the preferences of agents 3 and 4.

\noindent a) Say that agents 3 and 4 have the same preference $\succsim^a_i: 2,3,4,1,5,6$. To calculate $D(\succsim^a_{\{3,4\}},\succsim_{-\{3,4\}})$, first note that agents 3 and 4 both prefer agent 2 to staying single and agent 2 prefers 3 and 4 to staying single. So just like agents 1, 2, 5, and 6, agents 3 and 4 cannot be matched at Step 1 ($N_{stuck}^1=\emptyset$). Agents $2\in N^1_{up}$ and $3\in N^1_{down}$  are two $N^1$-adjacent singles, so $(2,3)\in \nu^2$. Without any other framed parties, Step 1 keeps all other agents single. At Step 2,
    $\nu^2$ currently partners agent 3 with her top choice 2.  Agent 3 is therefore  stuck, and  Step 2 finalizes the match $(2,3)\in D(\succsim^a_{\{3,4\}},\succsim_{-\{3,4\}})$. No other agent is stuck at Step 2, and $\nu^3$ is the restriction of $\nu^2$ to $\{1,4,5,6\}$. Agent 4 is stuck at $\nu^3$, and Step 3 finds the match $(4)$. Since no other agent is stuck at Step 3, $\nu^4$ is the restriction of $\nu^3$ to $\{1,5,6\}$, so $\nu^4=\{(1),(5),(6)\}$.
    At Step 4, $1\in N^3_{up} $ and $5\in N^3_{down}$ are two $\{1,5,6\}$-adjacent singles. Therefore Step 5 updates the current submatching to $\nu^5=\{(1,5),(6)\}$.
       Steps 5 and 6 then finalize these matches. We in sum get $D(\succsim^a_{\{3,4\}},\succsim_{-\{3,4\}})=\{(1,5),(2,3),(4),(6)\}$.

 \noindent b) Say that agent 3's preference is identical to that of agents 1 and 2: $\succsim^b_3: 6,5,4,3,2,1$, whereas agent 4's is identical to that of agents 5 and 6: $\succsim^b_4: 1,2,3,4,5,6$. As in a) no agent is stuck at Step 1. Step 1 updates the current submatching to $\nu^2=\{(1),(2),(3,4),(5),(6)\}$. No agent is stuck at Step 2.  At Step 2 there are no adjacent singles who prefer each other to being single. However the (non-empty) party, consisting of agents 3 and 4, is framed by the singles $2\in N^2_{up}$ and $5\in N^2_{down}$. So Step 2 updates the current submatching to $\{(1),(2,4),(3,5),(6)\}$. At Step 3 once again no agent is stuck. The party consisting of agents $\{2,3,4,5\}$ is framed by the singles $1\in N^3_{up}$ and $6\in N^3_{down}$. Step 3 then updates the current submatching to $\nu^4=\{(1,4),(2,5),(3,6)\}$. Since according to $\nu^4$  agents 3 and 4 date their top ranked partners, Step 4 finalizes the matches $(1,4)$ and $(3,6)$. Step 4 does not finalize the match $(2,5)$ since 2 and 6 as well at 1 and 5 prefer each other to their current partners. Step 5 then matches 2 with 5 and we get that $D(\succsim^b_{3,4},\succsim_{-\{3,4\}})=\{(1,4),(2,5),(3,6)\}$.

\noindent c) Say that according to $\succsim^c_{\{3,4\}}$ agents 3 and 4 top rank each other and rank themselves in second place. Step 1 of  case c) is then identical to
Step 1 of case b): the current submatching is updated to  $\nu^2=\{(1),(2),(3,4),(5),(6)\}$. Step 2 then finds that agents 3 and 4 are stuck with each other, so that $(3,4)\in D(\succsim^c_{\{3,4\}},\succsim_{-\{3,4\}})$. At Step 3 no agent is stuck. The currently single agents 2 and 5 are $\{1,2,5,6\}$-adjacent and prefer to date each other  to being single, so that $\nu^3=\{(1),(2,5),(6)\}$. At Step 4 no agent is stuck and the current submatching is updated to $\nu^4=\{(1,5),(2,6)\}$. Agents 5 and 6 are stuck at Step 5, and we conclude with $D(\succsim^c_{\{3,4\}},\succsim_{-\{3,4\}})=\{(1,5),(2,6),(3,4)\}$.

\section{The proof of Lemma \ref{the lemma}} \label{appendix: Lemma proof}

Fix a profile $\succsim\in \hat{\Omega}$.
To use induction note firstly that $\nu^1=id$ trivially implies that $\nu^{1}$ is made up of singles and parties and  that for any party $P$ at $\nu^1$: $L(P)\cap N^1_{down} =\emptyset= R(P)\cap N^1_{up}$.

\medskip

To prove Lemma \ref{figure: parties} it then suffices to show that for each  $k$ we have:

\medskip

$a[k]$: Step $k$ either matches or reassigns some agents. $b[k]$: $\nu^{k+1}$ is made up of singles and parties. $c[k]$: For any $i\in N^{k+1}$: $ \nu^{k+1}(i)\succsim_i \nu^k(i)$ and $(i-\ptop(\succsim_i))(\nu^k(i)-\nu^{k+1}(i))\geq 0$.
$d[k]$: for any party $P$ at $\nu^k$: $L(P)\cap N^{k+1}_{down} =\emptyset= R(P)\cap N^{k+1}_{up}$.

\medskip

For each Step $k$ partition $N^k$ into $N^k_{up}$, $N^k_{down}$ and $N^k_{stuck}$, the sets of agents that are upwardly mobile, downwardly mobile or stuck at $(\nu^k,\succsim)$.

\medskip

Observation $O$: $\nu^k(i)=\min N^k\Rightarrow i\notin N^k_{down}$, $\nu^k(i)=\max N^k\Rightarrow i\notin N^k_{up}$. The reason for the first statement is that an agent $i$ whose current partner is the smallest unmatched agent $\min N^k$ cannot prefer a yet smaller unmatched agent. There simply is no such smaller agent. Mutatis mutandis the second statement follows from the same reason.

\medskip

Step 1:

\medskip
 If $N_{stuck}^1\neq \emptyset$, Step 1  matches each $i\in N_{stuck}^1$. If $N_{stuck}^1=\emptyset$ then  Observation $O$ and the fact that each agent is single at $\nu^1$ imply that $\min N^1$ and $\max N^1$ are singles with
 $\min N^1\in N^1_{up} $ and  $\max N^1 \in N^1_{down} $. So  there exist at least two adjacent singles $j\in N^1_{up} $ and $j'\in N^1_{down}$ with $j<j'$. Update the current partners of  any two such agents  so that $(j,j')\in \nu^2$. So $a[1]$ holds. Since  any agent $i\in N^2$ is either single or paired with an adjacent agent, and since any two adjacent parties form a bigger party, $\nu^2$ is made of singles and parties and  $b[1]$ holds. Now fix any $i$ such that $\nu^2(i)\neq \nu^1(i)$, so that  $(i,j)\in \nu^2$  holds for $j\succ_i i$ and $j\in\{i+1,i-1\}$. If $j=i+1$, then $i\in N^1_{up} $ and $j\in N^1_{down}$, so that $j=\nu^2(i)\succ_i i=\nu^1(i)$. If $j=i-1$, then $i\in N^1_{down} $ and $j\in N^1_{up}$, so that $j=\nu^2(i)\succ_i i=\nu^1(i)$. In any case we get that $\nu^2(i)\neq \nu^1(i)\Rightarrow \nu^2(i)\succ_i \nu^1(i)$  and $(i-\ptop(\succsim_i))(\nu^1(i)-\nu^2(i))\geq 0$ and $c[1]$ holds. Now assume that $i\in L(P)$ for some party $P$ at $\nu^2$. Since all agents at $\nu^2$ are either single or matched with an adjacent agent, $\nu^2(i)=i+1$. By $c[1]$ $\nu^2(i)=i+1\succ_i \nu^1(i)=i$. Since agent $i$'s preference is single-peaked $i+1=\nu^2(i)\succ_i j$ for all $j<i+1$ so that $i\notin N^2_{down}$. We in sum see that $L(P)\cap N^2_{down} =\emptyset$. Applying the above arguments mutatis mutandis to $R(P)$, we see that also $R(P)\cap N^2_{up}=\emptyset$ holds. So in sum we get $d[1]$.

\medskip

Now assume $a[k-1]$, $b[k-1]$, $c[k-1]$ and $d[k-1]$ hold up to some $k\geq 2$.

\medskip

\noindent $a[k]$: Step $k$ either matches or reassigns some agents.

\medskip

If $N_{stuck}^k\neq \emptyset$, Step $k$ matches each $i\in N_{stuck}^k$. So assume that
 $N_{stuck}^k=\emptyset$. To see that  $\min N^k$   must be an  upwardly mobile single,
suppose $\min N^k$ was not single, so suppose that
  $i=\nu^k(\min N^k)>\min N^k$. Then
 $b[k-1]$ implies that  $i\in R(P)$ for some party $P$ at $\nu^k$ and $d[k-1]$ implies   $R(P)\cap N^k_{up} =\emptyset$. But Observation $O$ together with $N^k_{stuck}=\emptyset$ yield the contradiction $i\in N^k_{up}$. So $\min N^k$ must be single. Invoking once again Observation $O$ we see that
    $\min N^k\in N^k_{up}$. Mutatis mutandis the same arguments also imply that $\max N^k$ is a downwardly mobile single, $\nu(\max N^k)=\max N^k\in N^k_{down} $. Since $\nu^k$ is by $a[k-1]$ made up of singles and parties, at least one pair $l<r$ with $l\in N^k_{up}$ and $r\in N^k_{down}$ frames some party $P$.

 According to $\nu^{k+1}$ the agents in $L(P)\cup R(P)\cup \{l,r\}$ then form a new party $P'$ at $\nu^{k+1}$ with $L(P')\colon=\{l\}\cup L(P)$ and $R(P')\colon=\{r\}\cup R(P)$. To see that it is feasible to form such a party note firstly that  $L(P)\cup R(P)\cup \{l,r\}$ is a set of $N^k$-adjacent agents. Secondly  $\mid L(P)\mid =\mid R(P)\mid$
 implies that $\mid L(P')\mid =\mid \{l\}\cup L(P)\mid=\mid \{r\}\cup R(P)\mid=\mid R(P')\mid $. Finally $l<\min L(P)$, $\max R(P)<r$, and the assumption that $L(P),R(P)$ form the party $P$, mean that $i<\nu^{k+1}(i)$ holds for all $i\in L(P')$ if we match the agents in $L(P')$ with the agents in $R(P')$ according to their size (the smallest in $L(P')$ with the smallest in $R(P')$, the second smallest in $L(P')$ with the second smallest in $R(P')$, and so forth).
 We in sum see that Step $k$ either matches or reassigns some agents.

 \medskip

\noindent $b[k]$: $\nu^{k+1}$ is made up of singles and parties.

 \medskip

By $b[k-1]$ $\nu^k$ is made up of singles and parties.  Any two adjacent parties form a party. Moreover, if some but not all agents in a party are matched, the party members that remain unmatched continue to form a party. Finally if $\nu^{k+1}$ is obtained via reassignements from $\nu^k$, the instruction converts a framed party into a bigger party. So in all cases the current submatching $\nu^{k+1}$ derived in Step $k$ is once again made up of singles and parties.

 \medskip

\noindent $c[k]$: $i\in N^{k+1}\Rightarrow $: $\nu^{k+1}(i)\succsim_i \nu^k(i)$ and $(i-\ptop(\succsim_i))(\nu^k(i)-\nu^{k+1}(i))\geq 0$

\medskip

If $\nu^{k+1}(i)\neq \nu^k(i)$ then at $\nu^k$ there must exist a party $P$ that is framed by two singles $l<r$ with $l\in N^k_{up}$ and $r\in N^k_{down}$ such that
 $i\in \{l,r\}\cup L(P)\cup R(P)$.  Say Step $k$ updates the current partners of the agents in $\{l,r\}\cup L(P)\cup R(P)$ to a party $P'$ at $\nu^{k+1}$.
 First consider the case that $i\in L(P')=\{l\}\cup L(P)$. The assumptions that $l\in N^k_{up}$ and $N_{stuck}^k=\emptyset$ together with $d[k-1]$  imply  $L(P')\subseteq N^k_{up}$ and therefore $i\in N^k_{up}$. So for $i$ there exists some $j>\nu^k(i)$ such that $i\succ_j \nu^k(j)$ and $j\succ_i \nu^k(i)$. Since $\succsim_i$ is single-peaked $j'\succ_i \nu^k(i)$ then also holds for the smallest $j'>\nu^k(i)$ with $i\succ_{j'}\nu^k(i)$. According to the instructions for Step $k$, this agent $j'$ equals $\nu^{k+1}(i)$. We in sum get $\nu^{k+1}(i)\succ_i \nu^k(i)$   and  $(i-\ptop(\succsim_i))(\nu^k(i)-\nu^{k+1}(i))\geq 0$ for any $i\in L(P')$. Applying the same arguments mutatis mutandis to any $i\in R(P')$ we see that $\nu^{k+1}(i)\succ_i \nu^k(i)$ holds for all $i\in L(P')\cup R(P')$.
\medskip

\noindent $d[k]$: for any party $P$ at $\nu^k$: $L(P)\cap N^k_{down} =\emptyset= R(P)\cap N^k_{up}$.

\medskip
 Consider any $i\in L(P')$. By the preceding arguments $\nu^{k+1}(i)\succ_i \nu^k(i)$ holds for $\nu^{k+1}(i)$ the smallest agent $j'$ such that $j'>\nu^k(i)$ and $i\succ_{j'} \nu^k(j)$. The single peakedness of agent $i$'s preference together with $\nu^{k+1}(j)\succsim_j\nu^k(j)$ for all $j\in N^{k+1}$ then imply
  that $\nu^{k+1}(i)\succ_i j$ holds for any $j$ such that $j<\nu^{k+1}(i)$ and $i\succ_j \nu^{k+1}(j)$, so $i\notin N^{k+1}_{down}$, and consequently $L(P')\cap N^{k+1}_{down} =\emptyset$. Applying the same arguments mutatis mutandis to agents in $R(P')$, we see that also $R(P')\cap N^{k+1}_{up}=\emptyset$ holds.

\end{document}